\magnification = \magstep1
\baselineskip = 15pt
\def\vs{\vskip 10pt}
\centerline{\bf My Life with Fisher}
\font\ss = cmssbx10

\vs
\centerline{N.~David Mermin}
\centerline{Laboratory of Atomic and Solid State Physics}
\centerline{Cornell University, Ithaca, New York 14853-2501} 
\vs\vs
{\narrower\narrower

\baselineskip = 11pt
\noindent This is based on the after-dinner talk given at the 70th Birthday
Conference for Michael E.~Fisher at Rutgers in December, 2001.  It is
longer than the talk, incorporating additional text from the
after-dinner talk I gave at Fisher's 60th Birthday Conference at the
National Academy in Washington D.~C.~in 1991.

{}
}

\vs\vs

Many years ago I was writing a talk, ``My Life with Landau'', for a
conference commemorating the 80th anniversary of the birth of the
great L.~D.~Landau.  I knew I was going to have to deliver it before
an audience that included Michael Fisher, and I found to my distress,
as I sat there at the keyboard, that the image of Michael kept
intruding on my thoughts, questioning my assumptions, denouncing mean
field theories, and otherwise disrupting my concentration, in the way
that we have all come to know and love.  Finally, to chase him away, I
wrote ``Some day I would like to give a talk on `My Life with
Fisher'$\,$'' and strangely enough, that got rid of him.  But ever
since, I've known that the time would come when I would have to pay
for that liberating moment.

I first heard of Michael Fisher 38 years ago at the beginning of a
postdoctoral year at La Jolla.  I met another young postdoc, Bob
Griffiths, and in response to the intellectual sniffing out that goes
on at such occasions, Griffiths let it be known that what he was up to
was proving that the free energy of a spin system exists.  ``That it
{\it what?\/}'' I said.  ``That it {\it exists\/},'' said Griffiths
firmly.  ``I'm using some ideas I got from Michael Fisher.''  Well, I
thought, this Griffiths seems like a nice guy anyway.  And I decided
that this mentor of his, this Fisher, must be a man with deep
philosophical interests --- a sort of Plato of thermodynamics.

I didn't hear of Fisher again until I got to Cornell the next year and
Ben Widom told me one day that Michael Fisher was coming for a visit.
``That's nice'' I said, and remembering him as Griffiths' mentor,
looked forward to meeting such a quiet and contemplative man.  Well,
the visit lasted more than 20 years, and turned into by far the most
wonderful thing that has happened to me in my professional life.

Let me trace for you Michael's trajectory through the acknowledgments
sections of my publications.  He first shows up at the end of the 35
year old paper in which Herbert Wagner and I give our version of
Hohenberg's theorem.  Wagner and I had tried to explain to Michael
that an argument of Pierre's could be adapted to prove that there
could be no spontaneous magnetization in the 2-dimensional Heisenberg
model.  I hadn't known Michael for very long at that point, and one of
the first things I learned was that you should think twice before
claiming to {\it prove\/} something in front of a man who encourages
postdocs to show that the free energy {\it exists\/}.  He didn't
believe a word of it.  Spectral functions, indeed!  How did we know
those frequency integrals even converged?  It soon became evident that
we were dealing with a man who knew nothing about quantum field
theory, didn't care one bit that he didn't, and was convinced that we
would be better off ourselves to forget it.  Immediately.

So in the face of this astonishing attack, we worked backwards,
unbundling the result from the conceptual wrappings in which it was
enshrouded by some of the great thinkers of the previous decade,
peeling off layer after layer, day after day, in the face of
unrelenting skepticism, until finally we had it down to a trivial
statement about finite dimensional matrices.

And then an astonishing change took place.  ``Publish!'' he
practically shouted, ``it's very important!'' and having learned what
it was like to be at the end of a Michael Fisher attack, I suddenly
learned what it was like to have him on your side.  Freeman Dyson came
to town.  Michael introduced us.  ``Mermin and Wagner have 
proved that there's no spontaneous magnetization in the
2-dimensional Heisenberg model,'' Michael proudly informed him, as
Herbert and I basked in his admiration.  ``Of course there isn't.''
Dyson responded.  ``But they have {\it proved\/} that there isn't''
Michael insisted.  One Dyson eyebrow may have moved up half a
millimeter in response.  No matter.  I was hooked on arguing with
Michael Fisher.  My life would never be the same.

Here are some later acknowledgments:

In a 1967 footnote: ``The analysis given here was constructed at
the suggestion and with the vigorous assistance of M. E. Fisher.''
It's a footnote rather than an acknowledgment, because in those days
they wouldn't let you say anything human in an acknowledgment.

In 1968 we read: ``M. E. Fisher's insistence on the difficulty of
specifying a criterion for crystalline ordering led me to discard
several earlier versions of the argument.''

Skipping ahead to 1976: ``We are indebted to M. E. Fisher for lending
us what seems to be the only copy of de Gennes' book now in Ithaca.''

In 1977 we read: ``The importance of these considerations was
brought home to me by a ferocious lunchtime discussion with M. E. Fisher.''

In 1979: ``It was M. E. Fisher who first suggested and repeatedly
insisisted that I should publish my lecture notes, but I am not sure
he deserves thanks for this.''

Finally, in our solid state physics book, Neil Ashcroft and I, after
thanking 47 alphabetically arranged colleagues, devote a whole
paragraph to No.~48:
\vskip 5pt
{\narrower \narrower \baselineskip = 13pt

\noindent One person, however, has influenced almost every chapter.  Michael E.
Fisher, Horace White Professor of Chemistrry, Physics, {\it and\/}
Mathematics, friend and neighbor, gadfly and troubadour, began to read the
manuscript six years ago and has followed ever since, hard upon our tracks,
through chapter, and, on occasion, through revision and re-revision,
pouncing on obscurities, condemning dishonesties, decrying omissions,
labeling axes, correcting misspellings, redrawing figures, and often making
our lives very much more difficult by his unrelenting insistence that we
could be more literate, accurate, intelligible, and thorough.  We hope he
will be pleased at how many of his illegible red marginalia have found their
way into our text, and expect to be hearing from him about those that have
not.

}
\vskip 5pt
I call your attention to our characterization of Michael as a
gadfly. It was only after coming to know Michael that I fully
understood what the Athenians meant when they called Socrates a
gadfly, and shortly after that I also began to understand why they had
made him drink the hemlock.  I think most readers understood what we
meant by ``gadfly", until the book started being translated into other
languages.  It was Michael himself who reported to me, with only the
slightest tinge of acidity, that a Japanese friend had nervously asked
him why our preface called him a ``small, but loud and annoying
insect''.  

The Russian translator simply gave up and replaced ``gadfly'' with
``pedant''.  I knew the Polish translator had taken a more serious
approach to the problem, but I never got around to figuring out just
what it was that Michael was called in the Polish translation, until,
in preparing this 70th birthday speech, I sought help from Wojciech
Zurek:

{\narrower \narrower \baselineskip = 13pt
\vs

\noindent Dear Wojciech,

\vs
\noindent Could you help me with a translation?  In our book Neil
Ashcroft and I refer to Michael Fisher as `` gadfly and troubadour".
In the Polish edition ``gadfly and troubadour" comes out as {\it ciety
jak osa i wesoly jak trubadur\/}. My theory is that ``gadfly" has
become {\it ciety jak osa\/} and troubadour has been expanded to {\it
wesoly jak trubadur.\/} Am I right and can you give me a translation
of these phrases?  I have to give an after-dinner speech at a banquet
in Fisher's honor.

}

\noindent Here are some excerpts from Zurek's reply: 

{\narrower \narrower \baselineskip = 13pt
\vs
\noindent The translation is not bad,
though it does change the meaning of the original phrase a bit: {\it
ciety jak osa\/} means ``ready to bite like a wasp".  You could also
say {\it giez} (which is literal for ``gadfly"), but you would not say
this about anyone in an after dinner speech in his honor$\ldots\,.$
\vs  
\noindent On the other hand, {\it wesoly jak trubadur\/} (literally 
``gay as a troubadour") probably changes the intent. I am guessing
{\it wesoly\/} was added for reasons of symmetry, to balance the
{\it ciety}.
\vs 
\noindent  All the best,
\vs 
\noindent  Wojciech
 \vs
\noindent  P.S. Why are you giving your after dinner speach in Polish?

}
 
\noindent I replied as follows:

{\narrower \narrower \baselineskip = 13pt

\noindent Dear Wojciech,
\vs
\noindent You have persuaded me that Polish is too subtle a medium.  I will
speak in simple English.
\vs
\noindent Many thanks,
\vs
\noindent David
\vs
\noindent P.S.  You are right about {\it wesoly jak trubadur.\/}  We
had in mind Michael's fondness for travelling with his guitar.  Not
his disposition, in whatever sense of the word you prefer.

}

So much for Polish.  Earlier this year, in reassuring defiance of
all the reckless gossip about our book getting out of date, the first
German translation appeared.  Here Michael is our {\it Freund und
Nachbar, Troubadour und laestiger Zeitgenosse,\/}  so in certain
German circles, Michael is
now becoming known as a troublesome contemporary.
\vs
\vs

I was out of town for the great revolution of 1970-71. I spent that
academic year away from Ithaca, on leave in Rome, but Michael told me
all about it when I got home.  What particularly impressed me was
this: In the years before that {\it annus mirabilis\/} Ken Wilson
would drop by my office every year or two and and say mysterious
things about phase transitions.  When we were both 17 we had the same
German teacher as freshmen at Harvard, so I knew he was pretty smart,
but I really thought he was losing his marbles with this talk about
rolling balls up hill with just enough energy so they almost made it
all the way to the top.  And then all this sloppy stuff in momentum
space.  He didn't even know how to write proper integral signs.  So I
was really amazed to come back home and find that Michael --- a man
who was interested in whether the free energy {\it existed\/}, mind
you --- had just waded right in, and was even able to explain to me
what Ken had been trying to tell me.  He had even learned about
Feynman diagrams.

But in the middle of all that unrigorous slop, he never forgot about
his high standards.  He gave a wonderful colloquium on what
mathematical physics was all about.  This is a pretty hard thing to do
in a colloquium, but he managed to make it absolutely gripping.  I'd
just come up with my own definition of the difference between
mathematical physics and theoretical physics that I was planning to
use in a colloquium I was to give at Princeton the following week, so
I tried it out on Michael after his lecture: The distinction, I told
him, was not to be found in the physics, but in the sociology of
physics: theoretical physics was done by physicists who lacked the
necessary skills to do real experiments; mathematical physics was done
by mathematicians who lacked the necessary skills to do real
mathematics.  Michael was not amused.  ``I'd advise you not to say
that at Princeton,'' he snarled.  Well I did anyway, and it nearly set
off a riot.

He was right, but the nice thing about Michael is that he is always
ready to give you advice about anything whatsoever, and if you don't
take his advice, he doesn't hold it against you.  He never forgets, of
course, that you didn't, and is quite willing to remind you, very
sympathetically, when you get into trouble because you didn't.  The
reason he is so good at giving advice is that he thinks very seriously
about everything, and always seeks out the best advice himself.  He
once asked me how I would find out where to buy a typewriter in New
York city.  I said I really couldn't tell him, because all I would do
would be to ask my father-in-law.  ``What's his name?'' he asked.  The
next time I spoke to my father-in-law he remarked that a strange thing
had happened.  A man with a very loud voice had phoned him in his law
offices and asked where to buy a typewriter.  ``What did you do?'' I
asked.  ``I told him, of course,'' said my father-in-law impatiently.
He was like Michael in some ways.

We all know that Michael has strong opinions about everything, but
what always fascinates me about Michael's opinions is that although
they are the strongest and most forcibly argued opinions I have ever
encountered, I can never predict in advance what direction they will
point in.  Closely related to this is the most profound unwillingness
to settle for things the way they are that I have ever run across.

What does Michael Fisher do when he checks into a hotel room for a
night?  He rearranges the furniture.  He'll rotate the bed 90
degrees, put the TV in the closet to make more room on the desk, carry
the desk over to the window to get more light.  He is an inspiration
to me.  Often I find it valuable to ask myself at difficult moments,
what would Michael do?  This strategy is not to be confused with that
of the ``What Would Jesus Do?" movement, though a comparison can be
interesting.  Often the two questions can lead to quite different
answers.

Let me give you a recent example of the benefits of asking ``What
would Michael do?''  A few years ago I was at the annual meeting of
the Danish Physical Society which took place at a small conference
center south of Copenhagen.  Each conferee had a little apartment with
a tiny attic.  Downstairs was a living room and bathroom.  Up a narrow
ladder was a built in bed in a room with no light.  Since one used the
apartment only at night this was an irritating arrangement.  I don't
know how Jesus would have coped, but it was pretty clear to me what
Michael would have done.  So I dragged the mattress and bedding down
the ladder, remade the bed on the living room floor, and never climbed
up to the attic again. This solution would not have occurred to me if
I had not asked myself "What would Michael do?"

The next day various Danish conferees complained about the
arrangement.  Ah, I said, under such trying circumstances you should always
ask yourself what Michael Fisher would do.  That night the
air was filled with matresses hurtling down ladders.  I believe
there is now a flourishing "What Would Michael Do?" movement
among the Danish physicists.

Sometimes the answer to ``What would Michael do?'' is clear, but one
lacks the courage to do it.  Here is a good example:

Michael and I were flying from Copenhagen to Ithaca together.  The
flight stopped in London, but after we reboarded and the door had
shut, the plane was slow to leave the gate.  As time went on it began
to look more and more like we would miss the Ithaca flight.  When the
likelhood began to approach certainty, Michael, muttering that that
there was no reason to spend the night on a bench at Kennedy when he
had a brother-in-law in London, rose from his seat and announced to
the flight attendant that he was getting off.  ``You can't,'' she
said.  ``Yes I can,'' said he.  ``We're about to depart,'' she said.
``You've been saying that for an hour and a half,'' said he.
``Michael, sit down,'' I said.  ``Shut up,'' said he.  And he strode
past her toward the closed door.  ``Open the door and let me out,'' he
said in the general direction of the door.  ``Your baggage is on
board,'' said they.  ``Hold it for me in New York, I'll pick it up
tomorrow,'' said he.

And then something happened that I wouldn't have believed.  The door
opened, a ramp appeared, and shouting back to me (who had for some
time been pretending he was a complete stranger) ``See you tomorrow in
Ithaca!''  off he strode.  Immediately thereafter the door closed, and
the plane took off, landing in New York just in time for me to make
the Ithaca flight which had, as usual, been delayed. I got home
without any waiting at all.

I conclude the story of my life with Fisher with the tale of how
Dorothy and I came to own a microwave oven.  Six years ago I agreed to
spend three months in Leiden as Lorentz Professor.  My immediate
predecessor in that position was Michael E. Fisher.  I remarked to a
friend that Michael would be a tough act to follow.  No, he said, on
the contrary: following Michael had to be the easiest way to be
Lorentz Professor because, as he put it, ``Nothing you ask of them
will seem unreasonable.''

When we were first shown the Lorentz Professors' apartment, I was
surprised to see a microwave oven in the kitchen.  We had never had
one ourselves, so I remarked on what a well-appointed kitchen it was.
``Yes," our host said, ``the microwave is quite new.  We just got it
last year."  Apparently Michael, on first being shown the apartment,
had looked it over and said, ``What, no microwave?!''  So for three
months we enjoyed the Fisher microwave.  When we got home I looked
around our kitchen and said ``What, no microwave?!''  We have had one
ever since.

The Lorentz Professor sits at Lorentz's old desk.  Attached to it is a
brass plaque stating that between 1878 and 1912 the desk was used by
H. A. Lorentz.  At Lorentz's desk was a chair.  Attached to it I found
a brass plaque stating that in 1994 the chair was used by
M. E. Fisher.  Whatever Michael thought of H. A. Lorentz, he
apparently did not admire his notion of what made for an decent desk
chair.  As I result, I sat very comfortably for three months at the
Lorentz desk in the Fisher chair.  There cannot be many who, for so
long a period, have been made {\it more\/} comfortable by Michael.
Gadflies do not make people more comfortable.

I have to say that life in Ithaca without that kind of excitement is a
shadow of what it used to be.  Michael lived just down the street from
me.  A lot of physicists were in the neighborhood.  As you walked down
the street looking at the mailboxes you would read Berkleman, Mermin,
Widom,  {\ss  F\ I\ S\ H\ E\ R}, Webb.  On the other hand life in
Maryland seems to have heated up.  After Michael had moved there and
bought a new house, I asked how things were going.  ``Not well," he
said.  ``Why?'' I asked.  ``We decided to move the walls out 3 feet'',
he said.  ``Which walls?'' I asked.  ``All of them,'' he said.

I conclude this birthday speech as I began it, with another
acknowledgment.  This one is from my contribution to the Michael
Fisher 60th Birthday Festschrift ten years ago:

{\narrower \narrower

\noindent I would like to thank God for arranging our lives so I could spend
over two decades with Michael Fisher at Cornell, and His servant, the
National Science Foundation, for supporting this investigation through
Grant No. PHY9022796.

}
\noindent  I would be delighted to thank the National Science
Foundation for supporting this latest tribute to Michael Fisher under
Grant PHY0098429.  But I'm not sure God's servant would consider it an
appropriate use of His resources, so I won't.

\bye